\documentclass[conference]{IEEEtran}
\IEEEoverridecommandlockouts
\usepackage{cite}
\usepackage{amsmath,amssymb,amsfonts}
\usepackage{algorithmic}
\usepackage{graphicx}
\usepackage{textcomp}
\usepackage{xcolor}
\usepackage{multirow}
\usepackage{caption}
\usepackage{enumerate}
\usepackage{booktabs}
\usepackage{soul}
\usepackage{color, xcolor} 

\def\BibTeX{{\rm B\kern-.05em{\sc i\kern-.025em b}\kern-.08em
    T\kern-.1667em\lower.7ex\hbox{E}\kern-.125emX}}
\begin{document}

\title{Integrated Multi-Level Knowledge Distillation for Enhanced Speaker Verification \\
\thanks{This work was supported by National Key R\&D Program of China (No.2023YFB2603902) and Tianjin Science and Technology Program (No. 21JCZXJC00190).}
}

\author{
    \IEEEauthorblockN{
        Wenhao Yang\IEEEauthorrefmark{1}, 
        Jianguo Wei\IEEEauthorrefmark{1}, 
        Wenhuan Lu\IEEEauthorrefmark{1}, 
        Xugang Lu\IEEEauthorrefmark{4},
        Lei Li\IEEEauthorrefmark{2}\IEEEauthorrefmark{3}
    }
    \IEEEauthorblockA{
        \IEEEauthorrefmark{1}College of Intelligence and Computing, Tianjin University\\
        \IEEEauthorrefmark{4}National Institute of Information and Communications Technology, Japan\\
        \IEEEauthorrefmark{2}University of Washington 
        \IEEEauthorrefmark{3}University of Copenhagen\\
    }
}

\maketitle

\begin{abstract}

Knowledge distillation (KD) is widely used in audio tasks, such as speaker verification (SV), by transferring knowledge from a well-trained large model (the teacher) to a smaller, more compact model (the student) for efficiency and portability. Existing KD methods for SV often mirror those used in image processing, focusing on approximating predicted probabilities and hidden representations.
However, these methods fail to account for the multi-level temporal properties of speech audio. In this paper, we propose a novel KD method, i.e., Integrated Multi-level Knowledge Distillation (IML-KD), to transfer knowledge of various temporal-scale features of speech from a teacher model to a student model. In the IML-KD, temporal context information from the teacher model is integrated into novel Integrated Gradient-based input-sensitive representations from speech segments with various durations, and the student model is trained to infer these representations with multi-level alignment for the output. We conduct SV experiments on the VoxCeleb1 dataset to evaluate the proposed method. Experimental results demonstrate that IML-KD significantly enhances KD performance, reducing the Equal Error Rate (EER) by 5\%.

\end{abstract}


\section{Introduction}

Knowledge distillation transfers knowledge from one deep neural network (the teacher) to another (the student) \cite{2015Distilling,2015FitNets,2020Knowledge}. This approach aims to train a smaller, more efficient student model using knowledge from pre-trained models for deployment or other applications. Knowledge is typically transferred by minimizing the distance between the teacher's and student's outputs. For image and audio classification tasks, the objective function often involves Kullback-Leibler (KL) divergence for soft labels \cite{2015Distilling} or Euclidean distance for features \cite{2015FitNets}. 

Applying knowledge distillation from image tasks to audio tasks may omit task-specific speech information due to the unique temporal distributions of speech signals. In speaker verification, the speaker classification model serves as the embedding extractor, and KD is applied during the classification training stage. This process may compromise the alignment of speaker-specific or temporal position information \cite{wang2019knowledge, mingote2020knowledge}, particularly the long-short time scale distribution properties critical for capturing individual speaker characteristics. Similar phenomena have been observed in other tasks, potentially degrading the performance of the student model \cite{komodakis2017paying, cho2019efficacy, ojha2023knowledge}.

Deep neural networks for audio tasks often lack insight into how and where task-related information is captured within speech signals. For speaker verification, equipped with complex structures \cite{zhou2021resnext, desplanques2020ecapa} and numerous parameters \cite{chen2022wavlm, jung2022pushing}, these models can learn accurate speaker information from speech and achieve high performance on several benchmarks \cite{nagrani2017voxceleb, nagrani2020voxsrc, huh2023voxsrc}. However, even state-of-the-art models may fail to capture comprehensive speaker information and degrade under unseen data distributions \cite{vestman2020voice, 9914016}. Therefore, to retain as much task-related knowledge as possible, knowledge distillation in audio tasks like SV should aim to effectively transfer the complete dark knowledge of the teacher model.

Recent advances in knowledge distillation have highlighted that previous methods may overlook complex relationships between teacher and student model outputs. To retain as much task-related knowledge as possible, Contrastive Relation Distillation \cite{tiancontrastive} transfers contrastive relations between samples, while multi-level alignment, including instance, class, and batch levels of statistics, has shown superior performance \cite{jin2023multi}. Despite this progress, these approaches have not been extensively explored in audio tasks. While classification tasks in both image and audio domains share commonalities, significant differences exist. Audio signals are time-series data with variable lengths, and rescaling speech can alter speaker identities \cite{yamamoto2019speaker}. In contrast, images can be uniformly rescaled. Moreover, speaker information may be distributed over longer durations \cite{li2023explore}, whereas objects of interest in images are typically localized.

\begin{figure*}[tb]
\centering
\includegraphics[width=0.85\linewidth]{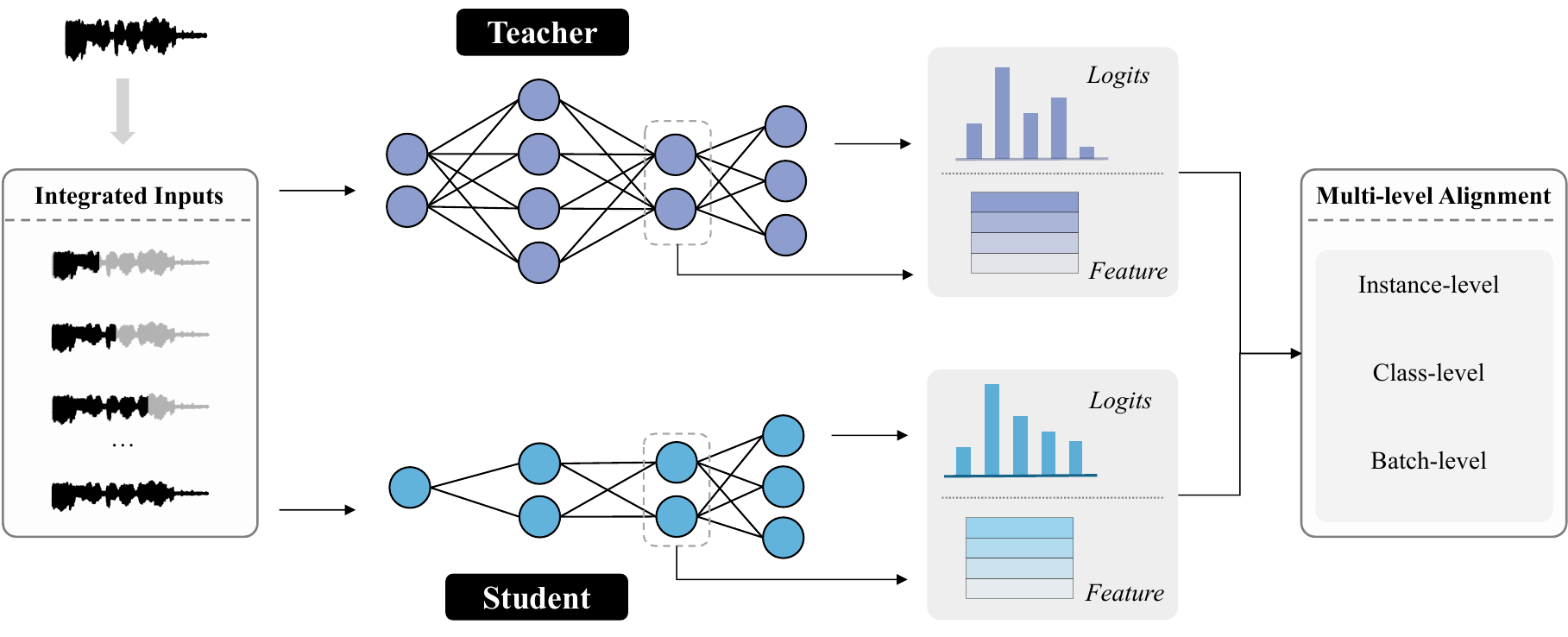}
\caption{Integrated Multi-level Knowledge Distillation for Speaker Verification.}
\label{fig:iml_kd}
\end{figure*}

Based on these discrepancies, we propose integrating interpretability research with knowledge distillation to interpret and transfer comprehensive speech context-specific knowledge for audio tasks. Several studies have focused on model interpretability for speaker verification. Li et al. \cite{li2022reliable} used Layer-CAM \cite{jiang2021layercam} to visualize relevant regions in acoustic features. Comparative analyses of class activation mapping methods, such as Grad-CAM \cite{selvaraju2017grad} and Score-CAM \cite{wang2020score}, have been conducted. Zhang et al. \cite{zhang2023study} applied Integrated Gradients (InteGrad) \cite{sundararajan2017axiomatic} and SHAP \cite{lundberg2017unified} to generate voiceprint maps, revealing speaker-discriminative information. InteGrad is particularly appealing due to its axiomatic attribution approach, which uses step-wise inputs to attribute the prediction of a deep network to its input features, thus elucidating model sensitivity and ensuring implementation invariance.

In this paper, we propose Integrated Multi-level Knowledge Distillation (IML-KD) for audio tasks. The Integrated Input module, derived from Integrated Gradients, interprets the model's sensitivity and implementation invariance to varying speech durations. The Multi-level Alignment module, based on Multi-level KD and embedding learning, transfers contextual speech task-related knowledge from the teacher to the student model. The contributions of this paper are as follows:

\begin{enumerate}
\itemsep=0pt
\item We propose utilizing the Integrated Gradients-based method to extract the audio task-related knowledge of the teacher model for knowledge distillation.
\item We introduce a multi-level and effective knowledge alignment method.
\item We evaluate the method on the speaker verification task and provide a comparative analysis of how knowledge distillation operates between teacher and student models.
\end{enumerate}

\section{Methodology}
\label{sec:format}

The Integrated Multi-level Knowledge Distillation contains two modules, including Integrated Input for model output attribution and multi-level knowledge distillation to align the output of the teacher and student models as in Figure~\ref{fig:iml_kd}.

\subsection{Preliminary}

\subsubsection{Integrated Gradient}

To find out what the neural network learned in classification, the saliency map could be seen as the attention map to which the model pays attention. The Gradient, as the easiest method to compute such a saliency map, could be extracted by back-propagation \cite{zeiler2014visualizing} because it shows how the output changes when the input changes. We regard the neural network as a complex non-linear function \( f(x) \), which represents the probability of speakers with input \(x\). The first-order Taylor expansion can approximate this complex function:

\begin{equation}
  f(x) \approx w^{T}x + b
  \label{eq3}
\end{equation}
where \(w\) is the derivative of \(f\) concerning the input \(x\) at the point \(x=x_0\):

\begin{equation} \label{eq4}
  L_{Gradient}(x) = w = \frac{\partial f(x)}{\partial x}  \bigg |_{x=x_0}
\end{equation}

\textbf{Integrate Gradient} \cite{sundararajan2017axiomatic} was proposed to attribute the output of deep neural networks like image models or text models to their input. InteGrad is an accumulated version of the Gradient method. It has been applied and proved to extract more accurate saliency maps. In SV, attributing the speaker's identity is close to finding out when and where the speaker's information is in speech signals. This is also the reason why we adopt this saliency map in our method.
\begin{equation}\label{eq:inte}
L_{InteGrad}(x) =  ( x - x' ) \times \frac{1}{m}\sum\limits_{k=1}^{m}\frac{\partial f( x' + \frac{k}{m}\times(x - x') )}{\partial x}
\end{equation}
where the m is the step size.

\subsubsection{Knowledge Distillation} 

In vanilla knowledge distillation (vanilla KD) \cite{2015Distilling}, the knowledge is transferred by the class probabilities produced by the teacher model as "soft targets" $y_t$. The final objective function of student models is the weighted sum on the cross entropy \(L_{ce}\) with the real labels \(l\) and KL divergence between predicted probabilities of the teacher \(y_t\) and the students $y_s$.

\begin{equation}
\begin{aligned}
L_{Vanilla} &= L_{ce}( y_s, l ) + \eta KL( y_t, y_s )
 \label{eq1}
 \end{aligned}
\end{equation}

In FitNets \cite{2015FitNets}, to minimize the difference in feature maps between the teacher and student, the distance \(D\) between these feature maps is added to the objective function. The embeddings (\( e_t, e_s\)) here could be speaker vectors. 

\begin{equation}
\begin{aligned}
L_{emb} = L_{ce}( y_s, l ) + \eta D( e_t, e_s ) 
\label{eq2}
\end{aligned}
\end{equation}

\subsection{Integrated Inputs Alignment}

To transfer the speech contextual modeling ability of models, we propose using Integrated Inputs. Specifically, we aim to ensure that the InteGrad for the input $x' + \frac{k}{m}\times(x - x')$ are aligned between the student and teacher. This can be achieved by aligning the outputs of step-wise inputs as defined in Equation.\ref{eq:inte}. By segmenting the input speech into patches along the time axis, each patch contains segments from different timestamps. Since audio classification task like speaker verification often relies on short utterances (2-4 seconds) extracted from longer ones, we propose dividing the longer utterance features into patches / segments for alignment. 

Using a predefined baseline, such as the average values of the entire input, it is easy to compute the Integrated Inputs. Given the step size $m$, the integrated input is equivalent to the Integrated Gradient:
 
\begin{equation}
\label{eq4}
 I_{Inte}(x) = \{ x' + \frac{k}{m} * (x - x') ) \}, k \in [1,m]
\end{equation}
where $x'$ is the baseline input and $m$ is the step size.

\subsection{Multi-level Knowledge Distillation}

After incorporating the speech contextual modeling methodology into Integrated Inputs for the teacher model, as proposed in \cite{jin2023multi}, we further align the multi-level outputs of the teacher model. Additionally, embeddings are included, as they play a crucial role in audio tasks such as speaker verification. The multi-level knowledge distillation encompasses three levels: instance-level, class-level, and batch-level knowledge.

For instance-level knowledge transferring, KL divergence is applied for the logits (prediction probabilities). Euclidean distance is applied for embeddings.

\begin{equation}\label{eq2}
\begin{split}
L_{inst} = KL(y_t, y_s) + D(e_t, e_s)
\end{split}
\end{equation}

For class-level and batch-level knowledge transferring, the Euclidean distance of the correlation matrix between classes or batch samples is considered. The embeddings are ignored for class-level distance computation.
\begin{equation}\label{eq2}
\begin{split}
L_{class} = \frac{1}{C} \sum \Vert Y_t - Y_s \Vert_2
\end{split}
\end{equation}
where $Y = y^Ty$ is a $C \times C$ matrix from prediction probability $y$ and $C$ is the number of classes. As embeddings are utilized finally and the output of the classifier is ignored in speaker verification, we add extra alignment for embeddings in batch-level alignment:

\begin{equation}\label{eq2}
\begin{split}
L_{batch} = \frac{1}{B} \sum ( \Vert Y'_t - Y'_s \Vert_2 + \Vert E_t - E_s \Vert_2 )
\end{split}
\end{equation}
where $Y' = yy^T$ is a $B \times B$ matrix, $B$ is the batch size and $E$ is the correlation matrix for embeddings.

The multi-level alignment KD loss can be expressed as : 

\begin{equation}
L_{ML} = L_{batch} + L_{class} + L_{inst}
\label{eq2}
\end{equation}

Finally, we unified the Integrated Inputs and Multi-level Alignment with classification training using CrossEntropy. The Integrated Multi-level KD can be accomplished with:

\begin{equation}\label{eq2}
L_{IML} = L_{ce}(x) + \eta L_{ML}(I_{inte}(x')) 
\end{equation}
where $x'$ is longer duration inputs and the weight $\eta$ is a constant knowledge distillation weight.

\section{Experiement}
\label{sec:format}

\subsection{Settings}

\noindent \textbf{Dataset} We use VoxCeleb1 \cite{nagrani2017voxceleb}, a large-scale English audio speaker dataset for our study. VoxCeleb1 has 1,211 speakers with 148,642 utterances for training and 40 speakers with 4,870 utterances for testing. 

\noindent \textbf{Model} ECAPA-TDNN \cite{desplanques2020ecapa} and ResNet \cite{zeinali2019but} are implemented within the PyTorch. The student model of ECAPA-TDNN is implemented with a channel number of 128 (ECAPA-C128). For model training, we use the same strategy as in \cite{desplanques2020ecapa}. Input features are 80-dimensional Mel-banks extracted from 2-second segments 
 for baselines, and 6-second segments for IML-KD. Data augmentation follows the method used in the training of \textit{x-vectors} \cite{snyder2018x}. We adopt the AAM-Softmax \cite{deng2019arcface} loss, with a scale $s$ of 30 and a margin $m$ of 0.2. 

\noindent \textbf{Implement details} The hyperparameter is selected as previous methods. The $\eta$ for $Embedding_{l2}$ and $Embedding_{cos}$ are 1 and 20 respectively. The $\eta$ for other methods is set to 9. The segment duration is 6 seconds for Integrated Inputs. During testing, cosine similarity is computed. Model performance is evaluated using the Equal Error Rate (EER) and Minimum Detection Cost Function (minDCF) \cite{martin2010nist}  where $p_{target}=0.01$ averaged over 3 random seeds, with both lower values indicating better model performance. 

\subsection{Results}

\begin{table}[h]
\centering
\caption{Performance of the knowledge distillation in speaker verification.}
\begin{tabular}{llcccc}
\toprule
 \multicolumn{1}{c}{\multirow{5}{*}{\textbf{Method}}} &   &  EER $\downarrow$ & minDCF $\downarrow$                                        & EER $\downarrow$  & minDCF $\downarrow$                                                                              \\   \cmidrule{2-6}

 & \multicolumn{1}{c|}{\multirow{2}{*}{teacher}} & \multicolumn{2}{c}{ECAPA}                                          & \multicolumn{2}{c}{ResNet34}                                                                                 \\
  &  \multicolumn{1}{c|}{} &                                                   2.66  & 0.2772 & 3.37   &  0.3397                                                                \\
  \cmidrule{2-6}
  & \multicolumn{1}{c|}{\multirow{2}{*}{student}} & \multicolumn{2}{c}{ECAPA-C128}                                          & \multicolumn{2}{c}{ResNet18}                                                                            \\
    & \multicolumn{1}{c|}{}  &                                            3.59 &  0.3672 &    3.76   &   0.3585                                                                   \\
\midrule
\multicolumn{2}{l}{KD \cite{2015Distilling}}  &  3.48 & 0.3544 & 4.01 & 0.3850 \\
\multicolumn{2}{l}{Embeddings$_{l2}$ \cite{mingote2020knowledge}}  &  3.29 & 0.3190 & 3.55 & 0.3508 \\
\multicolumn{2}{l}{Embeddings$_{cos}$ \cite{mingote2020knowledge}}  &  3.43 & 0.3589 & 3.63 & 0.3514 \\
\multicolumn{2}{l}{Multi-level \cite{jin2023multi}}  &  3.36 & 0.3350 & 4.06 & 0.3728 \\  
\multicolumn{2}{l}{\textbf{IML-KD (Ours)}}   &  \textbf{3.02} & \textbf{0.3169} & \textbf{3.51} & \textbf{0.3499} \\  
\bottomrule
\end{tabular}
\label{tab:overall}
\end{table}

\begin{figure}[htb]
\begin{minipage}[b]{\linewidth}
  \centering
  \centerline{\includegraphics[width=\linewidth]{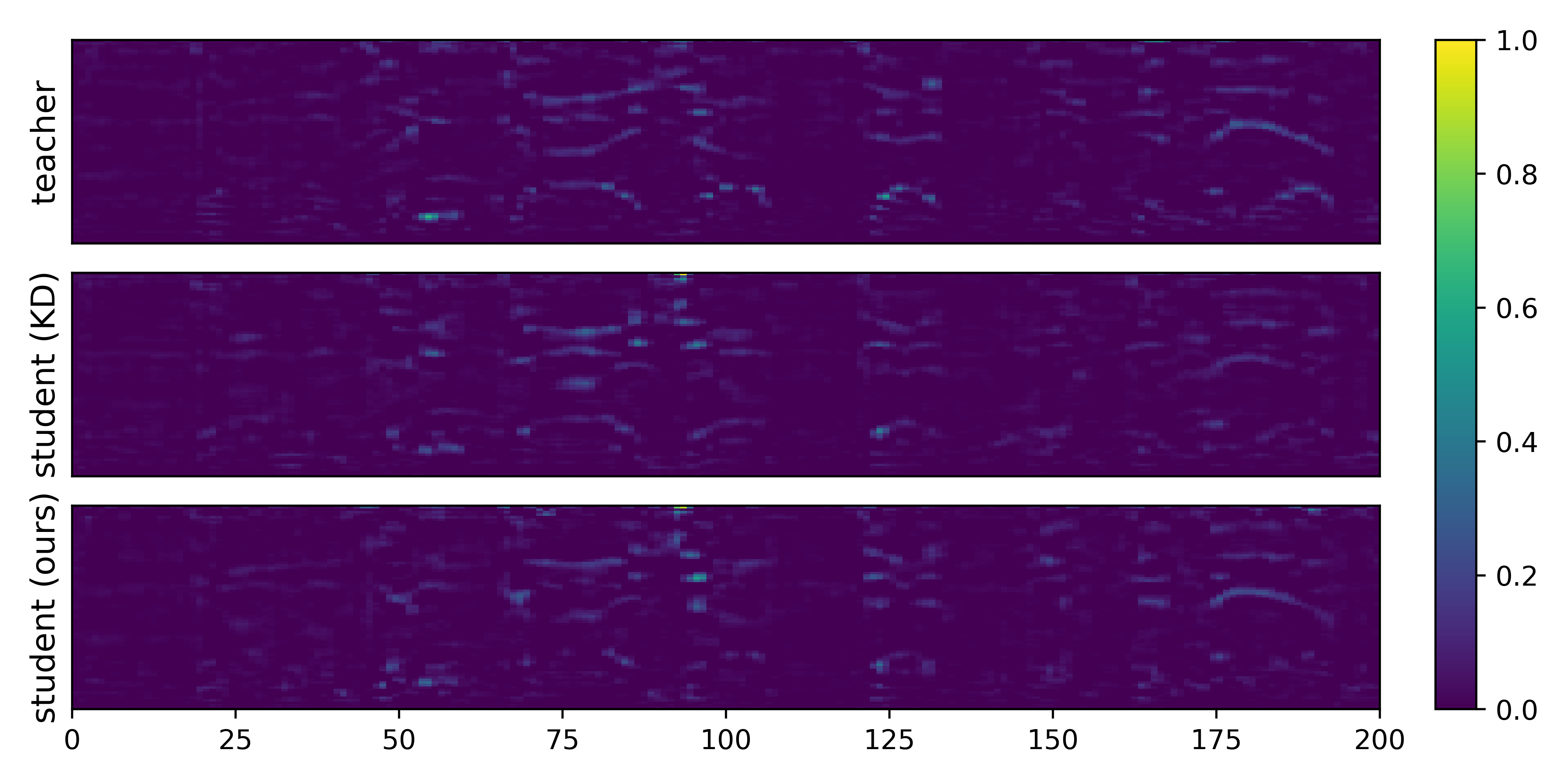}}
  \centerline{(a) Integrated Gradient for the Mel fbank.}\medskip
\end{minipage}
\hfill
\begin{minipage}[b]{0.92\linewidth}
  \centering
  \centerline{\includegraphics[width=\linewidth]{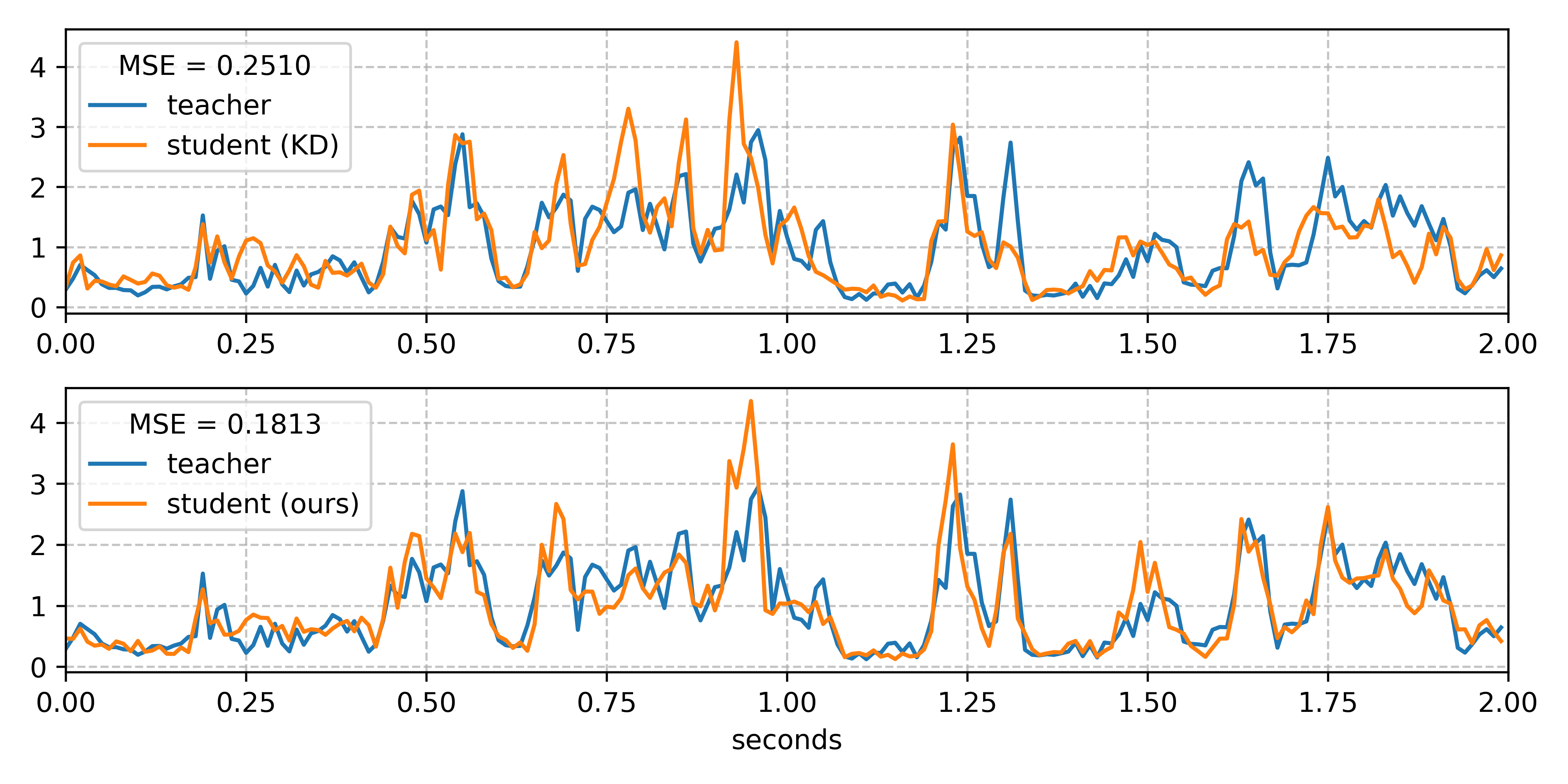}}
  \centerline{(b) Integrated Gradient weight along with time axis.}\medskip
\end{minipage}

\caption{ Visulization for utterance 'id10001-1zcIwhmdeo4-00001' in VoxCeleb1 dev.}
\label{fig:saliency}
\end{figure}

We present the overall results in Table~\ref{tab:overall}. We compare our method (IML-KD) with other KD methods on VoxCeleb1 test trials. The results show that our method achieves the lowest EER and MinDCF, with a particularly clear gap in EER. While some methods struggle with ResNet, our method consistently improves the performance of the student model (ResNet18).

We also plot Saliency Maps using InteGrad from the teacher (ECAPA-TDNN) model, KD, and IML-KD (ours) student models for speaker classification in Figure~\ref{fig:saliency}. It can be observed that the inference processes of the teacher and student models differ, as the attention regions in these saliency maps are not aligned. The student models are more concentrated on specific regions (around 1 second).

To be precise, the weight curves from the saliency maps along the time axis are also compared as shown in Figure~\ref{fig:saliency} (b). We compute the Mean Square Error (MSE) value for these curves. The MSE value between the teacher model and the IML-KD student is lower than that between the teacher and the KD student. Notably, from 1.5 to 2 seconds, the difference is significantly reduced by IML-KD. Therefore, we can conclude that our method, which utilizes Integrated Input and Multi-level alignment, distills a better student model by encouraging the student to mimic the teacher's inference process over time.

\subsection{Ablation Studies}

\subsubsection{IML-KD Components}

\begin{table}[h]
\centering
\caption{Performance of the student model (ECAPA-C128) for proposed IML-KD modules.}
\begin{tabular}{lcccc}
\toprule
\multicolumn{1}{c|}{\multirow{2}{*}{\textbf{Method}}} & \multirow{2}{*}{\textbf{Duration}}  & \multicolumn{2}{|c}{\textbf{Vox1-O }}       \\
\multicolumn{1}{c|}{}                               &         & \multicolumn{1}{|c}{EER $\downarrow$  } & \multicolumn{1}{c}{MinDCF $\downarrow$}  \\
\midrule
-                                     &    -   & 3.58                  & 0.3460                            \\
LMFT \cite{thienpondt2021idlab}                                     &    -   & 3.45                  & 0.3444                            \\
IML-KD (ours)                                  &    -   &  3.09 & 0.3300                            \\
\midrule
+ML  &  -  &  3.28 & 0.3439  \\
\cmidrule{2-4}
\multirow{5}{*}{++Inte\_Inputs} &  2s  &  3.40 & 0.3465   \\
 &  4s  &  3.23 & 0.3303   \\
  &  6s  &  3.09 & 0.3300   \\
 &  8s  &  3.07 & 0.3103   \\
 &  10s  &  \textbf{3.03} & \textbf{0.3085}   \\

\bottomrule
\end{tabular}
\label{tab:insert}
\end{table}

\begin{figure}[htb]
\begin{minipage}[b]{.48\linewidth}
  \centering
  \centerline{\includegraphics[width=\linewidth]{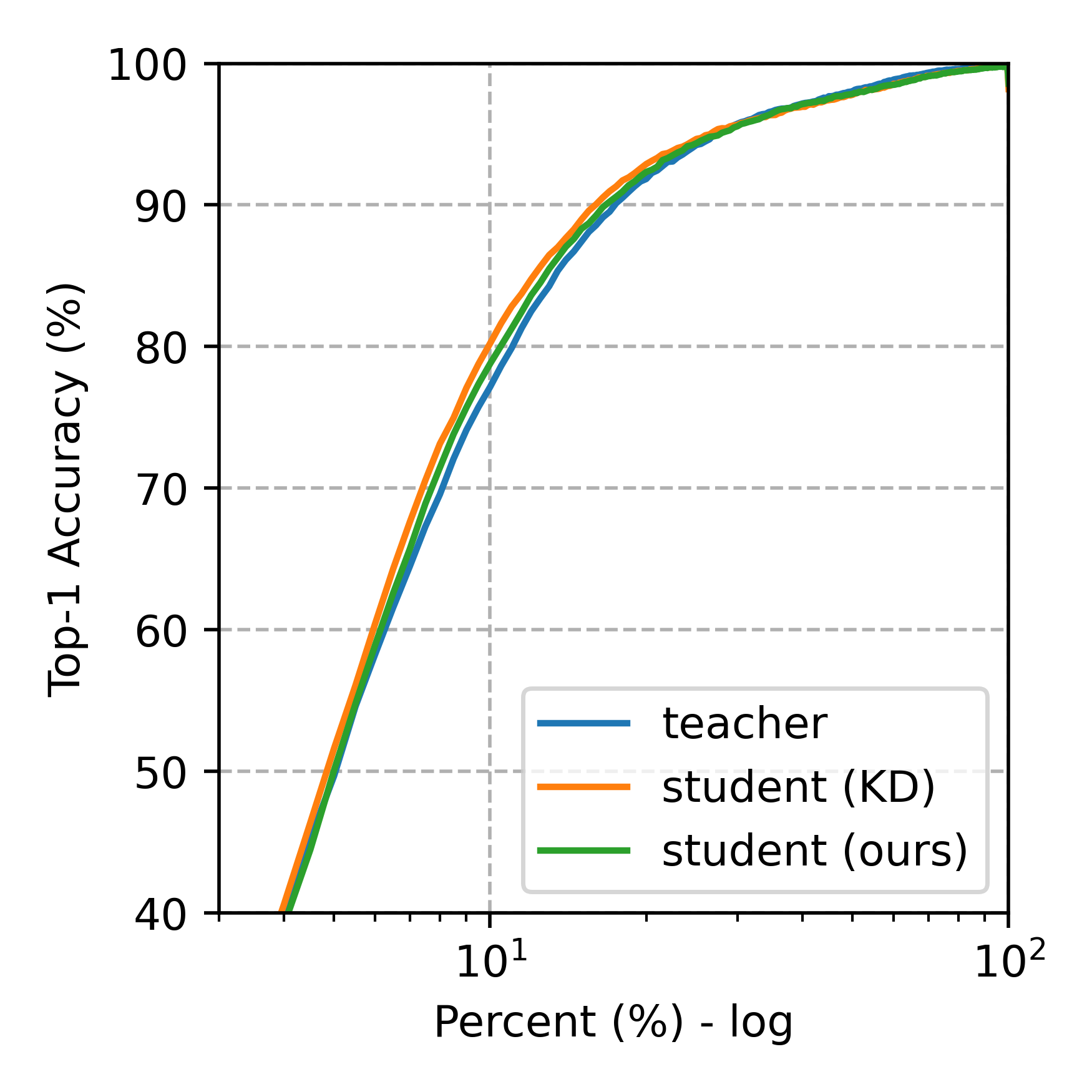}}
  \centerline{(a) Insertion }\medskip
\end{minipage}
\hfill
\begin{minipage}[b]{0.48\linewidth}
  \centering
  \centerline{\includegraphics[width=\linewidth]{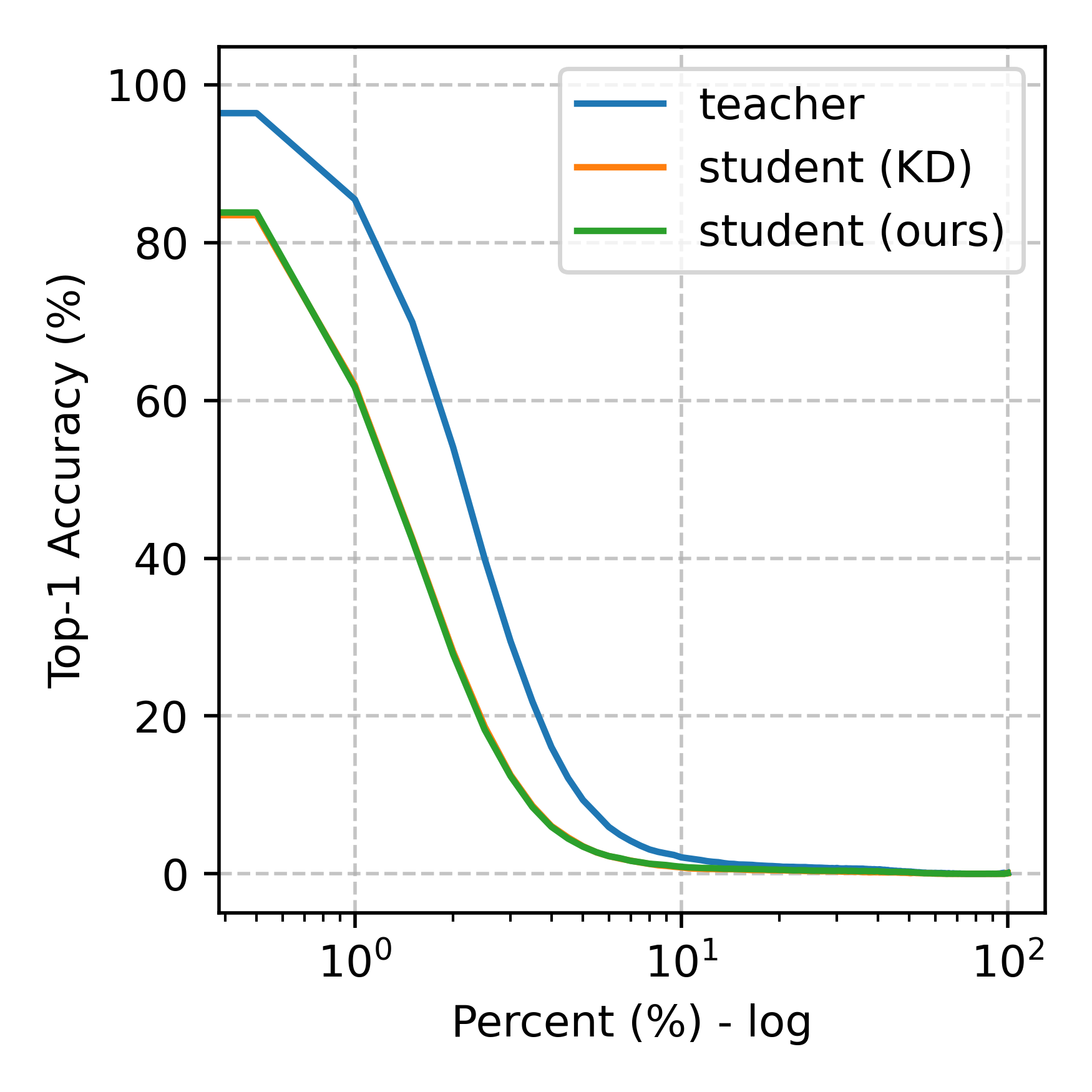}}
  \centerline{(b) Deletion }\medskip
\end{minipage}
\caption{ The deletion and insertion curve for student and teacher models.}
\label{fig:insert}
\end{figure}

We conducted ablation studies using ECAPA-C128 (student) to evaluate the IML-KD. The results are presented in Table~\ref{tab:insert}. It can be observed that both Integrated Inputs and multi-level alignment contribute to improved KD performance for speaker verification. Combining the two modules leads to a further reduction in EER. Additionally, compared with Large Margin Fine-tuning as in \cite{thienpondt2021idlab}, our approach demonstrates superior performance, implying that the improvement is not solely due to longer utterances.

Furthermore, the impact of Integrated Input duration was investigated in Table~\ref{tab:insert}. As the utterance duration increases, better alignment results for knowledge distillation are achieved. However, when the duration exceeds 6 seconds, the improvement becomes marginal.

\subsubsection{Deletion and Insertion Analysis}

We also investigate the behavior of student models in the speaker classification task using Insertion and Deletion analysis as described in \cite{petsiuk2018rise, li2022reliable}. The results are shown in Figure~\ref{fig:insert}.

It can be observed that the teacher model exhibits less accuracy drop during deletion and less accuracy gain during insertion compared to the student models. There is a noticeable gap between the student models and the teacher model. However, while there is almost no difference in the deletion curves in Figure~\ref{fig:insert} (b), our method (IML-KD) brings the insertion curve of the student model closer to that of the teacher model compared to vanilla KD in Figure~\ref{fig:insert} (a).

\section{Conclusion}
\label{sec:pagestyle}

In this paper, we propose a novel knowledge distillation method for audio tasks. Building on interpretability research in deep neural networks and recent advances in knowledge distillation, we introduce two modules: Integrated Inputs and multi-level alignment. Integrated Inputs extract task-related information comprehensively from the speech context, while multi-level alignment ensures this information is effectively transferred by aligning across samples, speakers, and batches.

Experiments on the speaker verification task using the VoxCeleb dataset demonstrate our method's effective distillation of speaker information over time, an aspect often overlooked by other methods. Visualization and analysis confirm the effectiveness and interpretability of our method, highlighting its potential to improve efficiency in other audio tasks.

\vfill\pagebreak

\bibliographystyle{IEEEtran}
\bibliography{refs}

\begin{thebibliography}{10}
\providecommand{\url}[1]{#1}
\csname url@samestyle\endcsname
\providecommand{\newblock}{\relax}
\providecommand{\bibinfo}[2]{#2}
\providecommand{\BIBentrySTDinterwordspacing}{\spaceskip=0pt\relax}
\providecommand{\BIBentryALTinterwordstretchfactor}{4}
\providecommand{\BIBentryALTinterwordspacing}{\spaceskip=\fontdimen2\font plus
\BIBentryALTinterwordstretchfactor\fontdimen3\font minus
  \fontdimen4\font\relax}
\providecommand{\BIBforeignlanguage}[2]{{%
\expandafter\ifx\csname l@#1\endcsname\relax
\typeout{** WARNING: IEEEtran.bst: No hyphenation pattern has been}%
\typeout{** loaded for the language `#1'. Using the pattern for}%
\typeout{** the default language instead.}%
\else
\language=\csname l@#1\endcsname
\fi
#2}}
\providecommand{\BIBdecl}{\relax}
\BIBdecl

\bibitem{2015Distilling}
G.~Hinton, O.~Vinyals, and J.~Dean, ``Distilling the knowledge in a neural
  network,'' \emph{Computer Science}, vol.~14, no.~7, pp. 38--39, 2015.

\bibitem{2015FitNets}
A.~Romero, N.~Ballas, S.~E. Kahou, A.~Chassang, C.~Gatta, and Y.~Bengio,
  ``Fitnets: Hints for thin deep nets,'' \emph{Computer ence}, 2015.

\bibitem{2020Knowledge}
J.~Gou, B.~Yu, S.~J. Maybank, and D.~Tao, ``Knowledge distillation: A survey,''
  2020.

\bibitem{wang2019knowledge}
S.~Wang, Y.~Yang, T.~Wang, Y.~Qian, and K.~Yu, ``Knowledge distillation for
  small foot-print deep speaker embedding,'' in \emph{ICASSP 2019-2019 IEEE
  International Conference on Acoustics, Speech and Signal Processing
  (ICASSP)}.\hskip 1em plus 0.5em minus 0.4em\relax IEEE, 2019, pp. 6021--6025.

\bibitem{mingote2020knowledge}
V.~Mingote, A.~Miguel, D.~Ribas, A.~Ortega, and E.~Lleida, ``Knowledge
  distillation and random erasing data augmentation for text-dependent speaker
  verification,'' in \emph{ICASSP 2020-2020 IEEE International Conference on
  Acoustics, Speech and Signal Processing (ICASSP)}.\hskip 1em plus 0.5em minus
  0.4em\relax IEEE, 2020, pp. 6824--6828.

\bibitem{komodakis2017paying}
N.~Komodakis and S.~Zagoruyko, ``Paying more attention to attention: improving
  the performance of convolutional neural networks via attention transfer,'' in
  \emph{ICLR}, 2017.

\bibitem{cho2019efficacy}
J.~H. Cho and B.~Hariharan, ``On the efficacy of knowledge distillation,'' in
  \emph{Proceedings of the IEEE/CVF international conference on computer
  vision}, 2019, pp. 4794--4802.

\bibitem{ojha2023knowledge}
U.~Ojha, Y.~Li, A.~Sundara~Rajan, Y.~Liang, and Y.~J. Lee, ``What knowledge
  gets distilled in knowledge distillation?'' \emph{Advances in Neural
  Information Processing Systems}, vol.~36, pp. 11\,037--11\,048, 2023.

\bibitem{zhou2021resnext}
T.~Zhou, Y.~Zhao, and J.~Wu, ``Resnext and res2net structures for speaker
  verification,'' in \emph{2021 IEEE Spoken Language Technology Workshop
  (SLT)}.\hskip 1em plus 0.5em minus 0.4em\relax IEEE, 2021, pp. 301--307.

\bibitem{desplanques2020ecapa}
B.~Desplanques, J.~Thienpondt, and K.~Demuynck, ``Ecapa-tdnn: Emphasized
  channel attention, propagation and aggregation in tdnn based speaker
  verification,'' \emph{arXiv preprint arXiv:2005.07143}, 2020.

\bibitem{chen2022wavlm}
S.~Chen, C.~Wang, Z.~Chen, Y.~Wu, S.~Liu, Z.~Chen, J.~Li, N.~Kanda,
  T.~Yoshioka, X.~Xiao \emph{et~al.}, ``Wavlm: Large-scale self-supervised
  pre-training for full stack speech processing,'' \emph{IEEE Journal of
  Selected Topics in Signal Processing}, vol.~16, no.~6, pp. 1505--1518, 2022.

\bibitem{jung2022pushing}
J.~W. Jung, Y.~J. Kim, H.~S. Heo, B.~J. Lee, Y.~Kwon, and J.~S. Chung,
  ``Pushing the limits of raw waveform speaker recognition,'' in
  \emph{Proceedings of the Annual Conference of the International Speech
  Communication Association, INTERSPEECH}, vol. 2022, 2022, pp. 2228--2232.

\bibitem{nagrani2017voxceleb}
A.~Nagrani, J.~S. Chung, and A.~Zisserman, ``Voxceleb: a large-scale speaker
  identification dataset,'' \emph{arXiv preprint arXiv:1706.08612}, 2017.

\bibitem{nagrani2020voxsrc}
A.~Nagrani, J.~S. Chung, J.~Huh, A.~Brown, E.~Coto, W.~Xie, M.~McLaren, D.~A.
  Reynolds, and A.~Zisserman, ``Voxsrc 2020: The second voxceleb speaker
  recognition challenge,'' \emph{arXiv preprint arXiv:2012.06867}, 2020.

\bibitem{huh2023voxsrc}
J.~Huh, A.~Brown, J.-w. Jung, J.~S. Chung, A.~Nagrani, D.~Garcia-Romero, and
  A.~Zisserman, ``Voxsrc 2022: The fourth voxceleb speaker recognition
  challenge,'' \emph{arXiv preprint arXiv:2302.10248}, 2023.

\bibitem{vestman2020voice}
V.~Vestman, T.~Kinnunen, R.~G. Hautam{\"a}ki, and M.~Sahidullah, ``Voice
  mimicry attacks assisted by automatic speaker verification,'' \emph{Computer
  Speech \& Language}, vol.~59, pp. 36--54, 2020.

\bibitem{9914016}
X.~Qin, D.~Cai, and M.~Li, ``Robust multi-channel far-field speaker
  verification under different in-domain data availability scenarios,''
  \emph{IEEE/ACM Transactions on Audio, Speech, and Language Processing},
  vol.~31, pp. 71--85, 2023.

\bibitem{tiancontrastive}
Y.~Tian, D.~Krishnan, and P.~Isola, ``Contrastive representation
  distillation,'' in \emph{International Conference on Learning
  Representations}, 2020.

\bibitem{jin2023multi}
Y.~Jin, J.~Wang, and D.~Lin, ``Multi-level logit distillation,'' in
  \emph{Proceedings of the IEEE/CVF Conference on Computer Vision and Pattern
  Recognition}, 2023, pp. 24\,276--24\,285.

\bibitem{yamamoto2019speaker}
H.~Yamamoto, K.~A. Lee, K.~Okabe, and T.~Koshinaka, ``Speaker augmentation and
  bandwidth extension for deep speaker embedding,'' 2019.

\bibitem{li2023explore}
Z.~Li, Z.~Zhao, W.~Wang, P.~Zhang, and Q.~Zhao, ``Explore long-range context
  features for speaker verification,'' \emph{Applied Sciences}, vol.~13, no.~3,
  p. 1340, 2023.

\bibitem{li2022reliable}
P.~Li, L.~Li, A.~Hamdulla, and D.~Wang, ``Reliable visualization for deep
  speaker recognition,'' \emph{arXiv preprint arXiv:2204.03852}, 2022.

\bibitem{jiang2021layercam}
P.-T. Jiang, C.-B. Zhang, Q.~Hou, M.-M. Cheng, and Y.~Wei, ``Layercam:
  Exploring hierarchical class activation maps for localization,'' \emph{IEEE
  Transactions on Image Processing}, vol.~30, pp. 5875--5888, 2021.

\bibitem{selvaraju2017grad}
R.~R. Selvaraju, M.~Cogswell, A.~Das, R.~Vedantam, D.~Parikh, and D.~Batra,
  ``Grad-cam: Visual explanations from deep networks via gradient-based
  localization,'' in \emph{Proceedings of the IEEE international conference on
  computer vision}, 2017, pp. 618--626.

\bibitem{wang2020score}
H.~Wang, Z.~Wang, M.~Du, F.~Yang, Z.~Zhang, S.~Ding, P.~Mardziel, and X.~Hu,
  ``Score-cam: Score-weighted visual explanations for convolutional neural
  networks,'' in \emph{Proceedings of the IEEE/CVF conference on computer
  vision and pattern recognition workshops}, 2020, pp. 24--25.

\bibitem{zhang2023study}
J.~Zhang, L.~He, X.~Guo, and J.~Ma, ``A study on visualization of voiceprint
  feature,'' in \emph{Proc. INTERSPEECH 2023}, 2023, pp. 2233--2237.

\bibitem{sundararajan2017axiomatic}
M.~Sundararajan, A.~Taly, and Q.~Yan, ``Axiomatic attribution for deep
  networks,'' in \emph{International conference on machine learning}.\hskip 1em
  plus 0.5em minus 0.4em\relax PMLR, 2017, pp. 3319--3328.

\bibitem{lundberg2017unified}
S.~Lundberg, ``A unified approach to interpreting model predictions,''
  \emph{arXiv preprint arXiv:1705.07874}, 2017.

\bibitem{zeiler2014visualizing}
M.~D. Zeiler and R.~Fergus, ``Visualizing and understanding convolutional
  networks,'' in \emph{European conference on computer vision}.\hskip 1em plus
  0.5em minus 0.4em\relax Springer, 2014, pp. 818--833.

\bibitem{zeinali2019but}
H.~Zeinali, S.~Wang, A.~Silnova, P.~Mat{\v{e}}jka, and O.~Plchot, ``But system
  description to voxceleb speaker recognition challenge 2019,'' \emph{arXiv
  preprint arXiv:1910.12592}, 2019.

\bibitem{snyder2018x}
D.~Snyder, D.~Garcia-Romero, G.~Sell, D.~Povey, and S.~Khudanpur, ``X-vectors:
  Robust dnn embeddings for speaker recognition,'' in \emph{2018 IEEE
  International Conference on Acoustics, Speech and Signal Processing
  (ICASSP)}.\hskip 1em plus 0.5em minus 0.4em\relax IEEE, 2018, pp. 5329--5333.

\bibitem{deng2019arcface}
J.~Deng, J.~Guo, N.~Xue, and S.~Zafeiriou, ``Arcface: Additive angular margin
  loss for deep face recognition,'' in \emph{Proceedings of the IEEE/CVF
  Conference on Computer Vision and Pattern Recognition}, 2019, pp. 4690--4699.

\bibitem{martin2010nist}
A.~F. Martin and C.~S. Greenberg, ``The nist 2010 speaker recognition
  evaluation.'' in \emph{Interspeech}, vol. 2010, 2010, p. 2726.

\bibitem{thienpondt2021idlab}
J.~Thienpondt, B.~Desplanques, and K.~Demuynck, ``The idlab voxsrc-20
  submission: Large margin fine-tuning and quality-aware score calibration in
  dnn based speaker verification,'' in \emph{ICASSP 2021-2021 IEEE
  International Conference on Acoustics, Speech and Signal Processing
  (ICASSP)}.\hskip 1em plus 0.5em minus 0.4em\relax IEEE, 2021, pp. 5814--5818.

\bibitem{petsiuk2018rise}
V.~Petsiuk, A.~Das, and K.~Saenko, ``Rise: Randomized input sampling for
  explanation of black-box models,'' \emph{arXiv preprint arXiv:1806.07421},
  2018.

\end{thebibliography}

\vspace{12pt}

\end{document}